\documentstyle[plett,preprint]{aps}
\begin{document}
\draft
%
%
\hoffset=0.0in
%

\newcommand {\omgrho}	{\mbox{0.34}}	
\newcommand {\dsetaprho}	{\mbox{0.92}}	
\newcommand {\etaprho}	{\mbox{0.24}}	
\newcommand {\hftrho}	{\mbox{0.27}}	
\newcommand {\kstrho}	{\mbox{0.16}}	
\newcommand {\phirho}	{\mbox{0.04}}	
%
\def\DP{\ifmmode D^+ \else $D^+$\fi}
\def\F{\ifmmode D_s^\pm \else $D_s^\pm$\fi}
\def\FP{\ifmmode D_s^+ \else $D_s^+$\fi}
\def\Mmin{\ifmmode M_{\rm min} \else $M_{\rm min}$\fi}
\def\kstar{\ifmmode K^{*0} \else $K^{*0}$\fi}
\def\kstarb{\ifmmode \overline{K}^{*0} \else $\overline{K}^{*0}$\fi}
\def\kstarbno{\ifmmode \overline{K}^{*} \else $\overline{K}^{*}$\fi}
\def\kstno{\ifmmode \overline{K}^{*}(892)^0 \else $\overline{K}^{*}(892)^0$\fi}
\def\munu{\ifmmode \mu^{+}\nu \else $\mu^{+}\nu$\fi}
\def\etap{\ifmmode \eta^{\prime} \else $\eta^{\prime}$\fi}
\def\phidk{\ifmmode \FP\rightarrow\phi\munu
			\else $\FP\rightarrow\phi\munu$\fi}
\def\dsetapdk{\ifmmode \FP\rightarrow\etap\munu
			\else $\FP\rightarrow\etap\munu$\fi}
\def\dsetadk{\ifmmode \FP\rightarrow\eta\munu
			\else $\FP\rightarrow\eta\munu$\fi}
\def\rhodk{\ifmmode \DP\rightarrow\rho^{0}\munu
			\else $\DP\rightarrow\rho^{0}\munu$\fi}
\def\rhonodk{\ifmmode \DP\rightarrow\rho(770)^{0}\munu
			\else $\DP\rightarrow\rho(770)^{0}\munu$\fi}
\def\omgdk{\ifmmode \DP\rightarrow\omega\munu
			\else $\DP\rightarrow\omega\munu$\fi}
\def\etadk{\ifmmode \DP\rightarrow\eta\munu
			\else $\DP\rightarrow\eta\munu$\fi}
\def\etapdk{\ifmmode \DP\rightarrow\etap\munu
			\else $\DP\rightarrow\etap\munu$\fi}
\def\kstdk{\ifmmode \DP\rightarrow\kstarb\munu
			\else $\DP\rightarrow\kstarb\munu$\fi}
\def\kpinrdk{\ifmmode \DP\rightarrow (K^{-}\pi^{+})_{NR}\,\munu
			\else $\DP\rightarrow (K^{-}\pi^{+})_{NR}\,\munu$\fi}
\def\kstnodk{\ifmmode \DP\rightarrow\kstno\munu
			\else $\DP\rightarrow\kstno\munu$\fi}
\def\vcdvcs{\ifmmode \mid\!\!V_{cd}/V_{cs}\!\!\mid^2
                        \else $\mid\!\!V_{cd}/V_{cs}\!\!\mid^2$\fi}
\def\vcdvbu{\ifmmode \mid\!\!V_{cd}/V_{bu}\!\!\mid^2
                        \else $\mid\!\!V_{cd}/V_{bu}\!\!\mid^2$\fi}
%
\def\sameauthors#1{\hbox to\textwidth{\hss\vrule height.3cm width0pt\relax%
#1\hss}}
\def\etal{{\it et al.}\rm}
%
%
\begin{title}
Observation of \rhonodk
\end{title}
{}~\\
\normalsize
\noindent K. Kodama$^{(1)}$,
N. Ushida$^{(1)}$,
A. Mokhtarani$^{(2),(a)}$,
V.S. Paolone$^{(2)}$,
J.T. Volk$^{(2),(a)}$,
J.O. Wilcox$^{(2),(b)}$,
P.M. Yager$^{(2)}$,
R.M. Edelstein$^{(3)}$,
A.P. Freyberger$^{(3),(c)}$,
D.B. Gibaut$^{(3),(d)}$
% ...
%
\begin{center}
(Fermilab E653 Collaboration)
\end{center}
\newpage
\small
{\em
\noindent
$^{(1)}$Aichi University of Education, Kariya 448, JAPAN
\\
\noindent
$^{(2)}$University of California (Davis), Davis, CA 95616, USA
\\
\noindent$^{(3)}$Carnegie-Mellon University, Pittsburgh, PA 15213, USA
\\
\noindent
$^{(4)}$Chonnam National University, Kwangju 500-757, KOREA
\\
\noindent
$^{(5)}$Fermi National Accelerator Laboratory, Batavia, IL 60510, USA
\\
\noindent
$^{(6)}$Gifu University, Gifu 501-11, JAPAN
\\
\noindent
$^{(7)}$Gyeongsang National University, Jinju 660-300, KOREA
\\
\noindent
$^{(8)}$Kinki University, Higashi-Osaka 577, JAPAN
\\
\noindent
$^{(9)}$Kobe University, Kobe 657, JAPAN
\\
\noindent
$^{(10)}$Korea University, Seoul 136-701, KOREA
\\
\noindent
$^{(11)}$Nagoya Institute of Technology, Nagoya 466, JAPAN
\\
\noindent
$^{(12)}$Nagoya University, Nagoya 464, JAPAN
\\
\noindent
$^{(13)}$The Ohio State University, Columbus, OH 43210, USA
\\
\noindent
$^{(14)}$Okayama University, Okayama 700, JAPAN
\\
\noindent
$^{(15)}$University of Oklahoma, Norman, OK 73019, USA
\\
\noindent
$^{(16)}$Osaka City University, Osaka 558, JAPAN
\\
\noindent
$^{(17)}$Science Education Institute of Osaka Prefecture, Osaka 558, JAPAN
\\
\noindent
$^{(18)}$Toho University, Funabashi 274, JAPAN
\\
\noindent
$^{(19)}$Utsunomiya University, Utsunomiya 321, JAPAN
\\
\noindent
$^{(20)}$Wonkwang University, Iri 570-749, KOREA}
\normalsize

\begin{abstract}
\vskip1.0in
We report the first observation of the Cabibbo disfavored
semileptonic decay mode \rhonodk,
and measure its decay rate relative to the Cabibbo favored mode $\kstnodk$
to be
$${\Gamma(\rhonodk) \over \Gamma(\kstnodk)} = 0.044 \,\,^{+0.031}_{-0.025}
\,\,(stat.)  \pm 0.014\,\,(sys.) .$$
The results are compared to theoretical predictions and to previous
experimental upper limits.

\end{abstract}
\bigskip
\narrowtext


\newpage

Measurement of the elements of the Cabibbo-Kobayashi-Maskawa (CKM) matrix
is an important goal of experimental high energy physics, and the
semileptonic decays of charm and beauty mesons play a central
role in this endeavor.  Although calculation of semileptonic decay
rates requires model input for the form factors involved, uncertainties
in the model are expected largely to cancel in ratios of decay rates
for three-body modes containing only vector mesons or only pseudoscalar
mesons in the final state.
The ratio $\vcdvcs$ can thus be extracted from $\Gamma(D\rightarrow
\kstarbno\,l\,\nu) / \Gamma(D\rightarrow\rho\,l\,\nu)$.
Similarly, $\vcdvbu$ can be obtained by comparing
$\Gamma(D\rightarrow\rho\,l\,\nu)$ and
$\Gamma(B\rightarrow\rho\,l\,\nu)$; Heavy Quark Effective Theory\cite{HQETIW}
predicts the uncertainty in this comparison to be small\cite{HQETDIB}.
Clearly, measurement of Cabibbo suppressed semileptonic
decay modes of the $D$ meson is required to obtain these ratios of
CKM matrix elements. To date, the only modes that
have been detected are\footnotemark
\footnotetext{Throughout this paper, charge conjugation  is implied unless
otherwise stated.}
$D^{0} \rightarrow \pi^{-} e^{+} \nu$\cite{MARK3DPIENU} and
$D^{+} \rightarrow \pi^{0} l^{+} \nu$\cite{ARNE}.
However, these two modes may have more significant contributions
from non-spectator diagrams\cite{ISBPI} than $D\rightarrow\rho\,l\,\nu$;
such diagrams
would pose additional complications in extracting the CKM matrix elements.

In this paper we present a signal for the decay mode $\rhonodk$,
which constitutes the first observation of Cabibbo disfavored semileptonic
charm decay involving a vector meson in the final state.
The decay rate for this mode is compared to that for $\kstnodk$, also
measured by this experiment.
The results are compared to previously measured upper limits and also to
theoretical predictions.


Data for this analysis were taken at Fermilab by experiment E653,  in which a
600 GeV/c $\pi^-$ beam was incident on a $4.92\,$g/cm$^2$ thick nuclear
emulsion
target\cite{NIM}.  The trigger required both a beam particle to interact in the
target and a muon to penetrate $3900\,$g/cm$^2$ of absorber; $8.2\times10^6$
events, corresponding to $2.5\times10^8$ interactions, were recorded in a
two-spectrometer detector. The upstream spectrometer, beginning 57$\,$mm
downstream of the target, consisted of 18 planes of silicon microstrip
detectors followed by a wide aperture dipole magnet and 55 drift chamber
planes.  A downstream muon spectrometer began after 1700$\,$g/cm$^2$ of
absorber and consisted of 12 drift chamber planes on each side of an iron
toroidal magnet. With the offline requirement that the muon momentum be greater
than 8$\,$GeV/c, the tracking information and momentum remeasurement of the
trigger muon in this system provided clear identification of the muon track in
the upstream spectrometer.

Events studied in this work are candidates for semimuonic charm decays to three
charged particles in the final state (three-prong topology).  For the
$\DP$ meson, such decays are dominated by the Cabibbo favored channel
$\kstnodk$, $\kstarb \rightarrow K^{-}\pi^{+}$.
Since charged hadron identification was not available, masses were
assigned to the two hadron tracks in all three-prong vertices
according to the decay mode under study.
Consequently, there is
potentially a large overlap of the dihadron invariant mass distribution from
$\kstdk$ and that from the decay of interest, $\rhonodk$.
The key variable used to isolate the $\rhodk$ signal is the
minimum parent mass, \Mmin,
which is the smallest mass of the parent particle
that will allow momentum to be conserved at the decay vertex\cite{E653KMUNU}.
$$\Mmin =
\sqrt{M_{vis}^{2}+p_{\perp}^{2}} + \sqrt{m_{\nu}^{2}+p_{\perp}^{2}},$$
where $M_{vis}$ is the invariant mass of the charged decay tracks,
$p_{\perp}$ is their momentum transverse to the parent direction,
and the neutrino mass $m_{\nu}=0$.
For a particular $\DP$ semimuonic decay mode involving no neutral hadrons, and
with the hadron masses correctly assigned,
the distribution in $\Mmin$ has a cusp at the $\DP$ mass.


The initial selection of candidates for \rhodk,
$\rho^{0} \rightarrow \pi^{+}\pi^{-}$ required a reconstructed three-prong
secondary vertex that included the trigger muon and was located downstream
from the primary interaction vertex.
The muon charge was required to be equal to the net
charge of the secondary vertex. To suppress background arising from hadronic
interactions in material, the secondary vertex was required to be more than
three standard deviations outside of dense material.
At this stage, the dominant
background was fake secondary vertices caused by reconstruction errors.  Such
background also dominated ``wrong-sign" vertices, for which the  charge of the
muon was opposite to that of the vertex.  In order to suppress this type of
background and to enhance charm selection in an unbiased manner, selection
criteria were determined by comparing their effect on the wrong-sign vertices
and on simulated $\rhodk$ decays.
The simulation was GEANT based and included all
known experimental effects\footnotemark
\footnotetext{ The Monte Carlo simulation
included charge deposition in the silicon detectors and pulse
time-over-threshold in the drift chambers. The validation of the simulation
with respect to observed distributions is discussed in \cite{JUDDTHES}. }.
The requirements most effective in suppressing reconstruction background were
that the $\chi^2$ of each track of the secondary vertex to belong to the
primary vertex be greater than 20,
that the decay length be greater than 5$\,$mm,
and that the momenta of the hadrons and the muon be greater than 5 and
20$\,$GeV/c, respectively.

Figure 1 shows the distribution in dipion invariant mass, $M_{\pi\pi}$,
obtained by assigning the pion mass to both hadronic tracks in secondary
vertices passing the initial selection criteria; the wrong sign vertices,
which are shown shaded, indicate the level of reconstruction background.
At this stage, before application of $\Mmin$ cuts,
the $M_{\pi\pi}$ distribution for the candidate charm decay vertices
is dominated by the peak due to the channel \kstdk,
$\kstarb \rightarrow K^{-}\pi^{+}$, for which the kaon was incorrectly
assigned the mass of a pion.

\marginpar{Figure 1 here}

The variable $\Mmin$ was utilized to suppress the mode $\kstdk$
explicitly\footnotemark
\footnotetext{Due to experimental resolution, the $\Mmin$ distribution extends
beyond the $\DP$ mass. One of the initial selection criteria was that the
minimum parent mass for the case of both hadronic tracks assigned the pion
mass, $\Mmin (\pi\pi\mu\nu)$, be greater than 1.6$\,$GeV/c$^2$ but less than
2.1$\,$GeV/c$^2$.}.
Figure 2 shows the distribution in minimum parent mass computed for a
$K^{-}\pi^{+}\mu^{+}\nu$ final state, $\Mmin(K\pi\mu\nu)$,
for simulated $\kstdk$ decays passing the initial selection criteria
(dotted histogram).
In the same figure, the $\Mmin (K\pi\mu\nu)$ distribution for simulated
$\rhodk$ decays is shown (dashed),
for which the hadronic track with charge opposite
that of the muon was assigned the mass of a kaon.
Most of the latter distribution is at higher values of $\Mmin (K\pi\mu\nu)$
than that for \kstdk, and
requiring $\Mmin(K\pi\mu\nu)$ to be greater than 2.0$\,$GeV/c$^2$ removes
95\% of the $\kstdk$ mode while retaining 67\% of the $\rhodk$ mode.
Figure 2 also shows the data (solid line) and the wrong-sign vertices (shaded).
Most of the data are removed by the cut on $\Mmin(K\pi\mu\nu)$, as expected,
since the distribution is dominated by \kstdk.
All of the wrong-sign vertices are removed by the cut.

\marginpar{Figure 2 here}

The reduction of $\kstdk$ yield by the  $\Mmin(K\pi\mu\nu)$ cut is
confirmed by
Figure 3, which shows the distribution in $K\pi$ invariant mass, $M_{K\pi}$,
for the data both before (solid line) and after (hatched) applying the cut.
The distribution for simulated $\rhodk$ decays surviving the cut
is also shown (dashed line).
Explicit exclusion of the $\kstarb$ mode from the data by further demanding
that $M_{K\pi}$ be less than 830$\,$MeV/c$^2$ or greater than 950$\,$MeV/c$^2$
removed  one additional event.

Figure 4 shows the $M_{\pi\pi}$ invariant mass distribution for the data
after application of all cuts.
(The one event removed by the requirements on $M_{K\pi}$ is shown hatched).
In this distribution, the $\rho^0$ signal region is defined by
$0.6 < M_{\pi\pi} < 0.9\,$GeV/c$^2$; the 6 data events within this
range constitute the final data sample.

\marginpar{Figure 3 here}
\marginpar{Figure 4 here}

Evidence that the events in the final data
sample are indeed charm decays is provided by the estimated
momenta and proper decay times of the parent particles\footnotemark
\footnotetext{For a $\pi^{+}\pi^{-}\munu$ final state, the
momentum of the parent
charm meson is unconstrained due to the unobserved neutrino,
and an estimate is required. The estimator used was that corresponding to the
most probable orientation of the neutrino momentum vector in the center of
mass of the charm meson\cite{JUDDTHES}.}.
The average momentum and transverse momentum of the candidate $\DP$ events
were
$157\pm 28\,$GeV/c and $1.40\pm0.19\,$GeV/c, respectively. These averages
agree well with
those for simulated $\rhodk$ decays, 144$\,$GeV/c and 1.14$\,$GeV/c, generated
with production parameters measured by this experiment\cite{JUDDXF}.
The average uncorrected proper decay time of the candidate events was
$1.43\pm0.30\,$ps, similarly in good agreement with that of the simulated
decays, 1.35$\,$ps.


In addition to the $\rhodk$ mode, there are potential contributions to the
final
data sample from reconstruction errors, hadronic interactions, and from other
charm decay channels. The elimination of all wrong-sign vertices by the
selection criteria indicates that the background from the first two
of these sources is small.
In particular, the requirement that the secondary vertex be located
at least three standard deviations outside of dense material strongly
suppresses background from hadronic interactions.
To confirm that the secondary interaction background was
small, the requirement on the secondary vertex position was relaxed to
the condition that the vertex only be outside of
dense material, while all other selection criteria were retained.
Although this procedure significantly increased the number of secondary
vertices, no evidence of a $\rho^0$ signal was observed in either the wrong
sign or correct sign data sets.
We therefore conclude that the contribution of
$\rho^0$ mesons from secondary interactions is
negligible in the final data sample.


The dominant background to $\rhodk$ is that from other decay modes of charm
mesons, some of which produce $\rho^0$'s indirectly.
The contribution from purely hadronic modes, in which
a daughter pion or kaon
decayed to a muon, was determined by Monte Carlo simulation and is
shown in Table I.
Several classes of semimuonic decays must be considered.
A few modes ($\phidk$ followed by $\phi\rightarrow K^{+}K^{-}$, $\etadk$
and \dsetadk) are effectively excluded by the $M_{\pi\pi}$ mass cuts.
The remaining background contributions from
Cabibbo favored three-body modes, Cabibbo favored four-body modes
and Cabibbo disfavored modes of the $\DP$ are treated separately in
the paragraphs below.
Where possible, backgrounds are determined from
experimental measurements; otherwise, predictions from the SI
model\footnotemark{
\footnotetext{The SI model\cite{DSTHESIS} is an improved version of the
ISGW quark potential model\cite{ISGW}.}
are used.


The backgrounds from the Cabibbo favored three-body channels $\kstdk$,
$\kstarb \rightarrow K^{-} \pi^{+}$ and $\phidk$, $\phi
\rightarrow \rho\pi$ or $\pi^{+}\pi^{-}\pi^{0}$ were determined from the
yields of $\kstdk$ and $\phidk$ measured in this
experiment\cite{E653BR,E653PHI}
and  from the efficiency of the particular mode to pass the selection
criteria\footnotemark
\footnotetext{All efficiencies were determined for $x_{F}>0$.}.
The expected contribution from $\dsetapdk$, $\etap \rightarrow \rho^{0}\gamma$
or $\omega\gamma$ was determined from the measured $\phidk$ yield using the
prediction of the SI model for $\Gamma(\dsetapdk) / \Gamma(\phidk)$
multiplied by
the factor of 1.49, which corrects for discrepancies between measured
and predicted decay rates\footnotemark
\footnotetext{As noted in the introduction,
 errors in the model are expected largely to cancel
in decay rate ratios for three-body modes involving only vector mesons, as
well as in ratios involving only pseudoscalar mesons.  The correction
factor of 1.49 is the ratio of the measured value of
$\Gamma(D\rightarrow K\,l\nu) / \Gamma(D\rightarrow\kstarb\,l\nu)$ to
the predicted value of the same quantity\cite{DOUGREV,E653PHI}.}.
Table I shows the number of events from these Cabibbo favored
channels contributing to
the final data sample and the corresponding efficiencies
determined by Monte Carlo simulation.

\marginpar{Table I here}

Resonant four-body semimuonic decay modes with a neutral hadron in the final
state, such as $D^{+} \rightarrow\overline{K}^{*0}\pi^{0}\mu^{+}\nu$ and
$D^{+} \rightarrow\overline{K}^{0}\rho^{0}\mu^{+}\nu$, result in a
three-prong secondary vertex and thus constitute a potential background.
The requirement that $\Mmin(K\pi\mu\nu)$ be greater than 2.0$\,$GeV/c$^2$
substantially suppresses any contribution from such modes, which populate
lower values of this variable due to the unobserved neutral hadron.
Measurements of branching ratios by this\cite{E6535P} and other
experiments\cite{E691RAPID} indicate a contribution from these channels
of less than 0.05 events at 90\% confidence level.

Nonresonant four-body modes such as $\kpinrdk$ are predicted to be
small relative to resonant modes\cite{ISGW}.
Measurements\cite{E691NR} confirm this prediction.
The mode $\DP\rightarrow (\pi^{+}\pi^{-})_{NR}\,\munu$ is assumed
to be similarly suppressed relative to the $\rhodk$ mode,
and the feedthrough into the $\rho^0$
signal region is further reduced by the $M_{\pi\pi}$ mass cuts.
Due to these considerations, both nonresonant and resonant four-body
semimuonic decay modes are regarded as negligible in the remainder of
this work.

The remaining background to consider was that from the Cabibbo disfavored
channels $\etapdk$ and $\omgdk$. After subtracting all the previously
discussed backgrounds, the final data sample consisted of
4.6 events. It  was assumed that these 4.6 events consisted only
of $\etapdk$, $\omgdk$ and \rhodk, and the relative contributions of these
three modes was computed from the SI model\footnotemark
\footnotetext{The prediction of $\omgdk$ relative to $\rhodk$ should be most
reliable since $\omega$ and $\rho$ differ only by isospin.}.
(The prediction for the $\etap$ mode was scaled by 1.49, as above.)
Table I shows the feedthrough from the $\etap$ and $\omega$ modes into
the $\rho^0$ calculated in this way.


Subtracting the sum of all background contributions from the final data
sample results in a signal of $4.0 \,\,^{+2.8}_{-2.3}$ $\rhodk$ events, for
which the efficiency is 2.12\%.
The simulated distribution in $M_{\pi\pi}$ for this mode and all charm
decay backgrounds, added with the appropriate efficiencies and branching
ratios, is shown superimposed on the data in Figure 4;
the simulation reproduces the observed spectrum well.
It is unlikely that the signal is a statistical fluctuation, since
the Poisson probability that the sum of backgrounds (2 events)
fluctuate up to or above the level of the observed candidates (6 events)
is 1.4\%.


To search for systematic effects,
the stability of the $\rhodk$ signal with respect to variations in the
selection criteria was studied.
Shifting the requirement on $\Mmin (K\pi\mu\nu)$ or on $\Mmin (\pi\pi\mu\nu)$
by $\pm 100\,$MeV/c$^2$ produced changes in yield for $\rhodk$ of less than
one half the statistical error, $\sigma$;
varying the $M_{K\pi}$ mass cuts by $\pm 20\,$MeV/c$^2$ at both ends
produced a change in yield of less than $0.25\sigma$;
expanding the allowed range in $M_{\pi\pi}$ to 0.5--1.0$\,$GeV/c$^2$,
or decreasing the range to 0.65--0.85$\,$GeV/c$^2$, resulted in less than a
$0.1\sigma$ change.
The systematic error on the $\rhodk$ yield  due to such variations is taken
to be $0.5\sigma$, so that the final signal is $4.0 \,\,^{+2.8}_{-2.3} \pm 1.3$
events.

A potential source of uncertainty in the $\rhodk$ yield is
due to the dominant background, \dsetapdk, which was estimated
with the SI model.
This experiment has searched for the decay
$\dsetapdk$, $\etap \rightarrow \eta\pi^{+}\pi^{-}$, $\eta \rightarrow
\pi^{+}\pi^{-}\pi^{0}$ explicitly\cite{E6535P}. One event was observed
in this five-prong topology, and an upper limit was obtained for $\dsetapdk$.
If the event is interpreted as an
observation of $\dsetapdk$, the feedthrough of this mode into the final data
sample would be $1.0\,\,^{+1.0}_{-0.8}$ event,
in agreement with the SI prediction shown in Table I.
This experiment\cite{E653PHI} has also measured
$\Gamma(D^{+}_{s} \rightarrow \eta \mu^{+}\nu + \etap \mu^{+}\nu)
/ \Gamma(\phidk) = 3.9 \pm 1.6$,
which is larger than, but marginally consistent with the scaled SI prediction
of 1.25.
With this experimental result, the measured $\FP$ yield and the
SI prediction that  $\Gamma(\FP\rightarrow\eta\,l^{+}\nu) /
\Gamma(\FP\rightarrow\etap\,l^{+}\nu) = 2.3$,
a contribution of $2.9 \pm 1.2$ background events
is estimated from the $\etap$ mode.  Consequently, we cannot rule out
experimentally the possibility that the $\rhodk$ signal has
a larger contribution from $\dsetapdk$ than estimated with the SI model.


{}From the yields and efficiencies for $\rhodk$ and $\kstdk$ measured by this
experiment, the following ratio of decay rates is determined:
$${\Gamma(\rhodk) \over \Gamma(\kstdk)} = 0.044 \,\,^{+0.031}_{-0.025}
\pm 0.014.$$
Using the world average decay rate\cite{PDG} for $\DP\rightarrow\kstarb\,l^{+}
\,\nu$, we obtain,
$$\Gamma(\rhodk) = (0.17 \,\,^{+0.12}_{-0.10} \pm 0.06) \times 10^{10}\,
{\rm s}^{-1}\, ,$$
which can be compared to the upper limit previously reported by the
Mark III experiment, $\Gamma(\DP\rightarrow\rho^{0}e^{+}\nu) <
0.35 \times 10^{10}\,{\rm s}^{-1}$ at 90\% confidence level\cite{MARK3UL}.

The SI model predicts the value of $\Gamma(\rhodk)/\Gamma(\kstdk)$
to be $0.46\,\mid\!\!V_{cd}/V_{cs}\!\!\mid^{2}$.  Since model
uncertainties are expected largely to cancel in this decay rate ratio,
the ratio of CKM matrix
elements $\mid\!\!V_{cd}/V_{cs}\!\!\mid^{2}$ can be extracted
by comparing the predicted and measured values of the decay rate
ratio.  However, the statistical precision of our observed
$\rhodk$ signal
precludes a precision measurement of the ratio of CKM matrix elements,
and it is more appropriate
to use the unitarity constraint on the CKM matrix\cite{PDG} in
a comparison of our
experimental result with the SI model.
With this additional input, the SI prediction gives
$\Gamma(\rhodk)/\Gamma(\kstdk) = 0.023$, which is consistent
with our measured value of $0.044 \,\,^{+0.031}_{-0.025} \pm 0.014$.


In conclusion, we have made the first observation of the Cabibbo disfavored
semimuonic decay \rhonodk, and measured its rate relative to \kstnodk.
Our measurement
for the ratio of decay rates $\Gamma(\rhonodk)/\Gamma(\kstnodk)$
agrees with the prediction of the SI quark potential model,
and confirms the expectation that Cabibbo disfavored decays
constitute a small
portion of the inclusive semileptonic decay rate of $D$ mesons.


\bigskip
We would like to thank D. Scora for making available a preliminary version
of his thesis, and N. Isgur for informative discussions.
We gratefully acknowledge the efforts of the Fermi National Accelerator
Laboratory staff in staging this experiment. This work was supported in
part by the US Department of Energy; the US National Science Foundation;
the Japan Society for the Promotion of Science; the Japan-US Cooperative
Research Program for High Energy Physics; the Ministry of Education,
Science and Culture of Japan; the Korea Science and Engineering Foundation;
and the Basic Science Research Institute Program, Ministry of Education,
Republic of Korea.

%
\newpage
%
\noindent{$^a$} Present Address: Fermilab, Batavia, IL 60510, USA.\\
$^b$ Present Address: Northeastern University, Boston, MA 02115, USA.\\
$^c$ Present Address: University of Florida, Gainsville, FL 32611, USA.\\
$^d$ Present Address: Virginia Polytechnic Institute and State University,
Blacksburg, VA 24061, USA.\\
$^e$ Present Address: Westinghouse Electric Corp., Pittsburgh, PA 15230, USA.\\
$^f$ Present Address: University of Iowa, Iowa City, IA 52242, USA.\\
$^g$ Present Address: SSC Laboratory, Dallas, TX 75237, USA.\\
$^h$ Present Address: University of California at Santa Barbara,
Santa Barbara, California 93106, USA.\\
$^i$ Present Address: University of Rochester, Rochester, NY 14627, USA.\\
$^j$ Present Address: Hirosaki University, Hirosaki-city 036, JAPAN.\\
$^k$ Present Address: Northern Illinois University, DeKalb, IL 60115, USA.\\
$^l$ Present Address: Soai University, Osaka 559, JAPAN. \\
$^m$ Present Address: University of Wisconsin, Madison, WI 53706, USA.\\
%
\newpage
\begin{table}
\caption{Backgrounds and efficiencies for final data sample.}
\begin{tabular}{|l|c|c|c|}
Decay mode & No. of events & Efficiency & Estimate source \\ \hline
 & & & \\
\dsetapdk & $\dsetaprho \pm 0.45$ & 0.24\% & SI
model\cite{DSTHESIS} and \\
\ \ \ \ \ \ \ \ \ $\etap \rightarrow \gamma\rho$ & & & E653 $\phidk$
yield\cite{E653PHI} \\
\ \ \ \ \ \ \ \ \ $\etap \rightarrow \gamma\omega$ & & & \\ \hline
 & & & \\
\kstdk & $\kstrho \,\,^{+0.13}_{-0.09}$ & 0.004\% & E653 $\kstdk$
yield\cite{E653BR}\\
\ \ \ \ \ \ \ \ \ $\kstarb \rightarrow K^{-} \pi^{+}$ & & & \\ \hline
 & & & \\
\phidk & $\phirho \,\,^{+0.04}_{-0.03}$ & 0.004\% & E653 $\phidk$ yield \\
\ \ \ \ \ \ \ \ \ $\phi \rightarrow \pi \rho$ & & & \\
\ \ \ \ \ \ \ \ \ $\phi \rightarrow \pi^{-}\pi^{+}\pi^{0}$ & & & \\ \hline
 & & & \\
\omgdk & $\omgrho \pm 0.20$ & 0.18\% & SI model \\
\ \ \ \ \ \ \ \ \ $\omega \rightarrow \pi^{-}\pi^{+}\pi^{0}$ & & & \\
\ \ \ \ \ \ \ \ \ $\omega \rightarrow \pi^{-}\pi^{+}$ & & & \\ \hline
 & & & \\
\etapdk & $\etaprho \pm 0.17$ & 0.36\% & SI model \\
\ \ \ \ \ \ \ \ \ $\etap \rightarrow \gamma\rho$ & & & \\
\ \ \ \ \ \ \ \ \ $\etap \rightarrow \gamma\omega$ & & & \\ \hline
 & & & \\
hadronic feedthrough & $\hftrho \pm 0.23 $ & & Simulation \\
\end{tabular}
\label{bckgnd}
\end{table}
%

\newpage
%
%
\figure{Distribution of $M_{\pi\pi}$ after applying the initial charm selection
criteria but before applying the $\Mmin (K\pi\mu\nu)$ cut.
The solid line histogram is the data, which is dominated
by a peak due to $\kstnodk$.
The distribution for the wrong-sign vertices is shown shaded.
\label{bigpipi}}
\figure{Distribution of $\Mmin(K\pi\mu\nu)$ after applying the initial charm
selection criteria; the data are indicated by the solid line histogram,
and the wrong-sign vertices are shown shaded.
The distribution for simulated $\kstdk$ is shown dotted and that for
$\rhodk$ dashed.
The arrow indicates the postion of the cut.
\label{kstmnmas}}
\figure{Distribution of $M_{K\pi}$ for the data (solid line histogram) after
applying the initial charm selection criteria.
The distributions after application of the $\Mmin(K\pi\mu\nu)$ cut are
shown for the data (hatched) and for simulated $\rhodk$ (dashed).
The range between the arrows indicates the region excluded by the $M_{K\pi}$
mass cuts.
\label{kstmass}}
\figure{Distribution of $M_{\pi\pi}$ after applying all selection criteria.
The data is the solid line histogram;
the one event removed by the $M_{K\pi}$ mass range cut is hatched.
The sum of simulated $\rhodk$ and charm decay backgrounds is shown dashed.
The arrows indicate the cuts used to define the $\rho^0$ mass region.
\label{rhomass}}
\end{document}